\documentclass[prd,superscriptaddress,unsortedaddress,twocolumn,preprintnumbers,amsmath,amssymb]{revtex4}
\usepackage[dvipdfmx]{graphicx}
\usepackage{amsmath,amssymb,times}
\usepackage{color}
\usepackage{ulem}
\usepackage{braket}
\usepackage{bm}
\usepackage{here}

\def\fun#1#2{\lower3.6pt\vbox{\baselineskip0pt\lineskip.9ptp
\ialign{$\mathsurround=0pt#1\hfil##\hfil$\crcr#2\crcr\sim\crcr}}}

\newcommand{\beq}{\begin{equation}}
\newcommand{\eeq}{\end{equation}}
\newcommand{\bea}{\begin{eqnarray}}
\newcommand{\eea}{\end{eqnarray}}

\DeclareSymbolFont{boldletters}{OML}{cmm} {b}{it}
\DeclareSymbolFontAlphabet{\mathbit}{boldletters}
\DeclareMathSymbol{\alpha}{\mathalpha}{letters}{"0B}
\DeclareMathSymbol{\beta}{\mathalpha}{letters}{"0C}
\DeclareMathSymbol{\gamma}{\mathalpha}{letters}{"0D}
\DeclareMathSymbol{\delta}{\mathalpha}{letters}{"0E}
\DeclareMathSymbol{\epsilon}{\mathalpha}{letters}{"0F}
\DeclareMathSymbol{\zeta}{\mathalpha}{letters}{"10}
\DeclareMathSymbol{\eta}{\mathalpha}{letters}{"11}
\DeclareMathSymbol{\theta}{\mathalpha}{letters}{"12}
\DeclareMathSymbol{\iota}{\mathalpha}{letters}{"13}
\DeclareMathSymbol{\kappa}{\mathalpha}{letters}{"14}
\DeclareMathSymbol{\lambda}{\mathalpha}{letters}{"15}
\DeclareMathSymbol{\mu}{\mathalpha}{letters}{"16}
\DeclareMathSymbol{\nu}{\mathalpha}{letters}{"17}
\DeclareMathSymbol{\xi}{\mathalpha}{letters}{"18}
\DeclareMathSymbol{\pi}{\mathalpha}{letters}{"19}
\DeclareMathSymbol{\rho}{\mathalpha}{letters}{"1A}
\DeclareMathSymbol{\sigma}{\mathalpha}{letters}{"1B}
\DeclareMathSymbol{\tau}{\mathalpha}{letters}{"1C}
\DeclareMathSymbol{\upsilon}{\mathalpha}{letters}{"1D}
\DeclareMathSymbol{\phi}{\mathalpha}{letters}{"1E}
\DeclareMathSymbol{\chi}{\mathalpha}{letters}{"1F}
\DeclareMathSymbol{\psi}{\mathalpha}{letters}{"20}
\DeclareMathSymbol{\omega}{\mathalpha}{letters}{"21}
\DeclareMathSymbol{\varepsilon}{\mathalpha}{letters}{"22}
\DeclareMathSymbol{\vartheta}{\mathalpha}{letters}{"23}
\DeclareMathSymbol{\varpi}{\mathalpha}{letters}{"24}
\DeclareMathSymbol{\varrho}{\mathalpha}{letters}{"25}
\DeclareMathSymbol{\varsigma}{\mathalpha}{letters}{"26}
\DeclareMathSymbol{\varphi}{\mathalpha}{letters}{"27}
\DeclareMathSymbol{\Gamma}{\mathalpha}{letters}{"00}
\DeclareMathSymbol{\Delta}{\mathalpha}{letters}{"01}
\DeclareMathSymbol{\Theta}{\mathalpha}{letters}{"02}
\DeclareMathSymbol{\Lambda}{\mathalpha}{letters}{"03}
\DeclareMathSymbol{\Xi}{\mathalpha}{letters}{"04}
\DeclareMathSymbol{\Pi}{\mathalpha}{letters}{"05}
\DeclareMathSymbol{\Sigma}{\mathalpha}{letters}{"06}
\DeclareMathSymbol{\Upsilon}{\mathalpha}{letters}{"07}
\DeclareMathSymbol{\Phi}{\mathalpha}{letters}{"08}
\DeclareMathSymbol{\Psi}{\mathalpha}{letters}{"09}
\DeclareMathSymbol{\Omega}{\mathalpha}{letters}{"0A}

\def\fun#1#2{\lower3.6pt\vbox{\baselineskip0pt\lineskip.9pt
\ialign{$\mathsurround=0pt#1\hfil##\hfil$\crcr#2\crcr\sim\crcr}}}

\begin{document}
\title{
A model consistent with LQCD data on $\rho$-meson screening mass
}

\author{Masahiro Ishii}
\email[]{masa1235@gmail.com}
\affiliation{Department of Physics, Graduate School of Sciences, Kyushu University,
             Fukuoka 819-0395, Japan}

\author{Akihisa Miyahara}
\email[]{miyahara94@gmail.com}
\affiliation{Department of Physics, Graduate School of Sciences, Kyushu University,
             Fukuoka 819-0395, Japan}

\author{Hiroaki Kouno}
\email[]{kounoh@cc.saga-u.ac.jp}
\affiliation{Department of Physics, Saga University,
             Saga 840-8502, Japan}  

\author{Masanobu Yahiro}
\email[]{orion093g@gmail.com}
\affiliation{Department of Physics, Graduate School of Sciences, Kyushu University,
             Fukuoka 819-0395, Japan}             

\date{\today}

\begin{abstract}
Recently, state-of-art LQCD calculations were done 
for  $\pi$-meson and $\rho$-meson screening mass, ${M_{\pi}^{\rm scr}}(T)$ and ${M_{\rho}^{\rm scr}}(T)$.
We consider the two-flavor system, and focus on temperature dependence $T$ of ${M_{\pi}^{\rm scr}}(T)$ and ${M_{\rho}^{\rm scr}}(T)$. 
Our aim is to construct a model consistent with LQCD data on 
${M_{\rho}^{\rm scr}}(T)$ and ${M_{\pi}^{\rm scr}}(T)$. 
\end{abstract}
\maketitle

\section{Introduction}
\label{Introduction}

Recently, state-of-art LQCD calculations were done 
for  $\pi$-meson and $\rho$-meson screening mass, ${M_{\pi}^{\rm scr}}(T)$ and ${M_{\rho}^{\rm scr}}(T)$,
 in finite temperature $T$~\cite{Cheng:2010fe,Maezawa:2016pwo}. 
In the present paper, we then concentrate on  ${M_{\pi}^{\rm scr}}(T)$ and its spin partner ${M_{\rho}^{\rm scr}}(T)$.

Meson masses can be classified into ``meson pole mass'' and 
``meson screening mass''. 
In LQCD simulations at finite $T$, 
meson pole (screening) masses are 
calculated from the exponential decay of temporal (spatial) 
mesonic correlation functions. 
LQCD simulations are more difficult for pole masses 
than for screening masses, since 
the lattice size is smaller in the time direction than 
in the spatial direction. This situation becomes more serious 
with respect to increasing $T$. 
For this reason, meson screening Masses have been calculated 
in most of LQCD simulations.

Effective models are an approach complementary to 
LQCD simulations. 
In fact, $T$ dependence of~$\rho$-meson pole mass was analyzed 
with the effective chiral theory~\cite{Song:1993af}, 
but the results are limited below the critical temperature $T_c$. 
When NJL-type models are used, 
$T$ dependence of $\pi$- and $\rho$-meson pole masses 
can be analyzed not only for $T<T_{c}$ but also for $T \ge T_{c}$, 
In fact, the $T$ dependence was investigated with the NJL model~\cite{He:1997gn,Blaschke:2001yj}.
As far as we know, there is no paper on  $T$ dependence of ${M_{\rho}^{\rm scr}}(T)$.

In general, the NJL model treats the chiral symmetry breaking, but not 
the deconfinement transition.  
Meanwhile, the Polyakov-loop extended Nambu--Jona-Lasinio (PNJL) model
~\cite{Meisinger,Dumitru,Fukushima1,Costa:2005,Ghos,Megias,Ratti1,Ciminale,Ratti2,Rossner,Hansen,Sasaki-C,Schaefer,Kashiwa1,Sakai1,Sakai2,Sakai_JPhys,Costa:2009,Ruivo:2012a}  and 
the entanglement PNJL (EPNJL) model~\cite{Sakai:2010rp,Sasaki_EPNJL} can deal with both  
the chiral symmetry breaking and the deconfinement transition. 
In the two-flavor case, LQCD shows that the chiral and deconfinement 
transitions take place simultaneously. The property can be explained 
not by the PNJL model but by the EPNJL model~\cite{Sakai:2010rp,Sasaki_EPNJL}. 
In the NJL-type models, it is difficult to calculate meson screening masses, since the calculation is time consuming~\cite{Florkowski}. This difficulty was solved by Ishii {\it et al.}~\cite{Ishii:2013kaa,Ishii:2015ira,Ishii:2016dln,Ishii:2018vvc}.  
As far as we know, there is no paper on ${M_{\rho}^{\rm scr}}(T)$ in the framework of the NJL and PNJL models. 

In this paper, we consider the two-flavor case with no chemical potential, and 
focus on $\pi$, $\rho$ mesons only. 
Our aim is to construct a model consistent with LQCD data on ${M_{\rho}^{\rm scr}}(T)$ and ${M_{\pi}^{\rm scr}}(T)$. 

\begin{figure}[htb]
\begin{center}
  \includegraphics[width=0.45\textwidth]{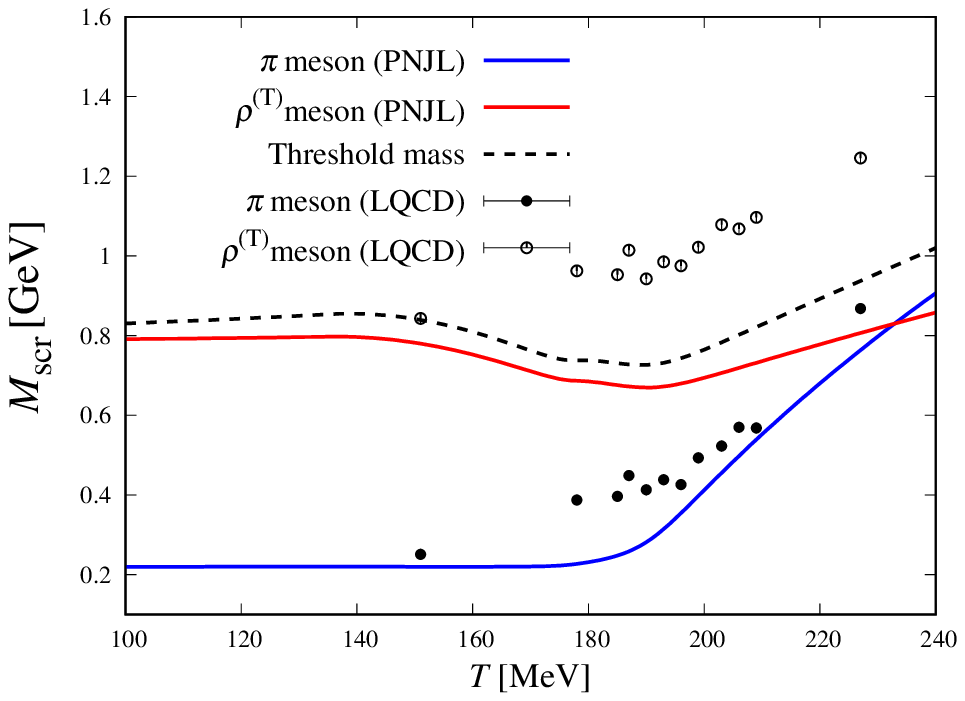}
  \includegraphics[width=0.45\textwidth]{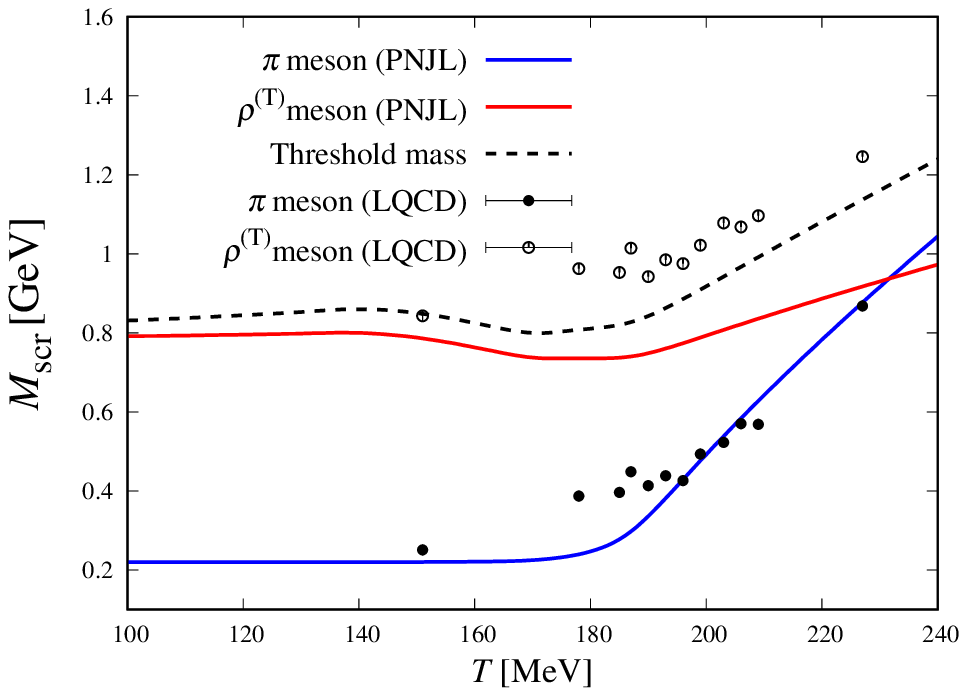}
\end{center}
\caption{
$T$ dependence of ${M_{\pi}^{\rm scr}}(T)$ and ${M_{\pi}^{\rm scr}}(T)$. 
LQCD data (dots) are taken  from Refs.~\cite{Cheng:2010fe,Maezawa:2016pwo}. 
In the PNJL model, the Polyakov-loop potential taken is logarithmic-type in the upper panel, 
but polynomial-type in the lower panel. 
The vector coupling $G_{\rm V}(T)$ is independent of $T$.
Note that $M_\pi^{\rm scr}(T)=M_\pi^{\rm pole}(T)$ at $T=0$. 
In the LQCD simulations, the $M_\pi^{\rm scr}(0)=220$~MeV is slightly heavier 
than the physical one ($M_\pi^{\rm pole}(0)=140$ MeV). 
We then change quark mass from $m_0=3.5$ MeV to $m_0=8.85$ MeV 
so as to reproduce $M_\pi^{\rm scr}(0)=220$ MeV. 
When we solve Eq.~\eqref{SD_rho_A1} for ${M_{\rho}^{\rm scr}}(T)$,  
the resulting ${M_{\rho}^{\rm scr}}(T)$ should be 
below the threshold mass $\mathcal{M}_{\rm lowest}$; 
see Eq.~\eqref{threshold}. 
The threshold mass (black-dots line) is always far 
below LQCD data on ${M_{\rho}^{\rm scr}}(T)$, indicating that  the PNJL model does not 
reproduce the LQCD data.  
  } 
\label{pi_rho_Polyakov-0}
\end{figure}

As shown in  the upper panel  of Fig.~\ref{pi_rho_Polyakov-0}, the PNJL model does not reproduce LQCD data~\cite{Cheng:2010fe,Maezawa:2016pwo} on ${M_{\rho}^{\rm scr}}(T)$ 
above $T_{\rm c}^{\chi}\simeq T_{\rm c}^{\rm d} \approx173$ MeV, when we take
the logarithm-type Polyakov-loop potential $\mathcal{U}$ of Ref.~\cite{Rossner}. 
In the lower panel, we take the polynomial-type  $\mathcal{U}$ (Poly-I) of Ref.~\cite{Haas:2013qwp}. 
The ${M_{\rho}^{\rm scr}}(T)$ for the polynomial-type  $\mathcal{U}$ is better agreement with the corresponding 
LQCD data than that for the logarithm-type Polyakov-loop  $\mathcal{U}$. 
From now on, we take the polynomial-type  $\mathcal{U}$. 
The polynomial-type  $\mathcal{U}$ reproduces the LQCD data for ${M_{\pi}^{\rm scr}}(T)$.  
Whenever we consider  $\pi$, the mixing between $\pi$ and $A_1$  is taken into account. 
Finally we consider magnetic-gluon contribution on ${M_{\rho}^{\rm scr}}(T)$ and ${M_{\rho}^{\rm scr}}(T)$.  
The results are consistent with the LQCD data for  ${M_{\rho}^{\rm scr}}(T)$ in $T > T_{\rm c}^{\rm d}$ and 
for  ${M_{\pi}^{\rm scr}}(T)$ in both $T > T_{\rm c}^{\rm d}$ and $T < T_{\rm c}^{\rm d}$. 
We call the present version of PNJL model ``magnetic-gluon  (MG) PNJL''.

The MG-PNJL  model is shown in Sec. \ref{sec:MG-PNJL} and 
numerical results are  in Sec. \ref{Numerical Results}. 
Section \ref{Summary} is devoted to a summary.


\section{MG-PNJL model}
\label{sec:MG-PNJL}

In order to construct the MG-PNJL  model, we consider the two-flavor case, 
since we  focus on $\pi$, $\rho$ mesons in the case of finite $T$.   

We start with the PNJL Lagrangian density with $T$-dependent  scalar four-quark coupling $G_{\rm S}(T)$ 
and a constant vector coupling $G_{\rm V}(0)$: Namely,  
\begin{eqnarray}
 {\cal L}  
&=& {\bar \psi}(i \gamma_\mu D^\mu -m_0)\psi  
+ G_{\rm S}(T)
[({\bar \psi}\psi )^2
 +({\bar \psi }i\gamma_5\bm{\tau}\psi )^2]
\nonumber\\
&+&G_{\rm V}(0)
\left[
\left(\bar{\psi}\gamma^{\mu}\bm{\tau}\psi
 \right)^{2} 
+ 
\left(\bar{\psi}\gamma^{\mu}\gamma_5\bm{\tau}\psi
 \right)^{2}
\right]
-{\cal U}(\Phi,{\bar \Phi},T),  
\nonumber\\
\label{L}
\end{eqnarray}
where $\psi$ is the quark field with the current quark mass $m_0$,  
${\vec \tau}$ stands for the isospin matrix. The covariant derivative $D^\nu$ is approximated into 
$\partial^\nu + \delta^{\nu}_{0}A^0_a{\lambda_a/2}$, where  the time component 
$A_{a}^{0}$ of the gauge field is treated 
as a homogeneous and static background field governed by the Polyakov-loop potential~$\mathcal{U}$. 
For $T$ dependence of $G_{\rm S}(T)$,  we assume   
\begin{eqnarray}
  G_{\rm S}(T)
   =
   \left\{
     \begin{array}{ll}
 G_{\rm S}(0) 
& (T < T_{\rm S}) \\
 G_{\rm S}(0) e^{-(T-T_{\rm S})^2/b_{\rm S}^2}
& (T \ge T_{\rm S}) \\
     \end{array}
     \right. .
 \label{T-dependent-Gs}
 \end{eqnarray}

As a Polyakov-loop potential $\mathcal{U}$, 
we consider two-types of $\mathcal{U}$. 
One is the logarithm-type potential of Ref.~\cite{Rossner}:
\bea
\frac{\mathcal{U}_{\rm log}(\Phi,\bar{\Phi},T)}
{T^4}
&=&
-\frac{a(T)}{2}\Phi\bar{\Phi} 
+ b(T)
{\rm ln}
\left[1-6\Phi\bar{\Phi} 
\right.\nonumber\\
&&\left. + 4\left(\Phi^3 + \bar{\Phi}^3\right)
-3(\Phi\bar{\Phi})^2
\right]
\label{log_ppot}
\eea
with 
\beq
a(T)=a_0 + a_1\left(\frac{T_0}{T}\right) + a_2\left(\frac{T_0}{T}\right)^2,~~
b(T) = b_3\left(\frac{T_0}{T}\right)^3
\eeq
and another is the polynomial-type potential of Ref.~\cite{Haas:2013qwp}:
\beq
\frac{\mathcal{U}_{\rm poly}(\Phi,\bar{\Phi},T)}{T^4}
=-\frac{b_2(T)}{2}\Phi\bar{\Phi}  
- \frac{b_3}{6}(\Phi^3 + \bar{\Phi}^3) 
+ \frac{b_4}{4}(\Phi\bar{\Phi} )^2
\label{pol_ppot}
\eeq
with 
\beq
b_2(T)=c_0 + c_1\left(\frac{T_0}{T}\right) + c_2\left(\frac{T_0}{T}\right)^2 + c_2\left(\frac{T_0}{T}\right)^3
\eeq
The parameters for each potential have been determined so as to reproduce 
thermodynamic quantities calculated with LQCD simulation in pure Yang--Mills theory. 
Their resultant values are summarized in Table~\ref{parameters_ppot}. 
These potentials have one dimensionful parameter $T_0$. The value is $T_0=270$ MeV in pure Yang--Mills theory. 
Once one considers quark degree of freedom, 
the parameter $T_0$ should be shifted to a lower value in association with change of typical energy scale 
through the QCD running coupling $g$.  
Hence we treat $T_0$ as an adjustable parameter and determine 
to be consistent with full QCD data on the chiral temperature 
$T_{\rm c} \simeq 173$~MeV with 10\% error~\cite{Karsch4}. 
The parameter thus obtained is $T_0=215$~MeV for each $\mathcal{U}$. 
\begin{table}[h]
\begin{center}
\caption
{Parameters taken in Polyakov-loop potentials}

\begin{tabular}{lccccccc}
\hline\hline
&$a_0$
&$a_1$
&$a_2$
&$b_3$
\\
\hline
Logarithm-type
&3.51
&-2.47
&15.2
&-1,75
\\
&$b_3$
&$b_4$
&$c_0$
&$c_1$
&$c_2$
&$c_3$
\\
Polynomial-type
&13.34
&14.88
&1.53
&0.96
&-2.3
&-2.85
\\
\hline
\end{tabular}
 \label{parameters_ppot}
\end{center}
\end{table}

In the Polyakov gauge, the Polyakov-loop $\Phi$ and 
its conjugate ${\bar \Phi}$ are obtained by  
\begin{align}
\Phi &= {1\over{3}}{\rm tr}_{\rm c}(L),
~~~~~{\bar \Phi} ={1\over{3}}{\rm tr}_{\rm c}({L^*})
\label{Polyakov}
\end{align}
with 
$L= \exp[i A_4/T]$ and 
$A_4/T={\rm diag}(\phi_1,\phi_2,\phi_3)$ satisfying the condition that ${\phi}_{1}+{\phi}_{2}+{\phi}_{3}=0$. 
It is then possible to choice ${\phi}_{3}=0$ and determine the  others  from ${\phi}_{3}=0$. This leads to
\bea
\phi_{2}=-\phi_{1}=\cos^{-1}{\left(\frac{3\Phi -1}{2}\right)} .
\label{eq:phi-Phi}
\eea

Making the mean field approximation (MFA), one can get 
the MFA Lagrangian density as 
\begin{align}
{\cal L}_{\rm MFA} ={\bar \psi}S^{-1}\psi 
- U_{\rm M}(\sigma) - {\cal U}(\Phi,{\bar \Phi},T)  
\label{eq:E7} 
\end{align}
with the quark propagator 
\begin{align}
S =\frac{1}{ i \gamma_\mu D^\mu- M }
\label{eq:E7} 
\end{align}
for 
\begin{eqnarray}
M = m_0 - 2G_{\rm S}(T)\sigma,~
U_{\rm M}= G_{\rm S}(T) \sigma^2,~
\sigma = \langle \bar{\psi}\psi \rangle.
\label{eq:E10}
\end{eqnarray}
We then obtain the thermodynamic potential $\Omega$ as 
\begin{equation}
\Omega = U_{\rm M}+{\cal U} + \Omega_{\rm F}
\label{PNJL-Omega} 
\end{equation}
with   
 \begin{align}
 \Omega_{\rm F} &= -2 N_{\rm c} N_{\rm f}\int \frac{d^3{\bm p}}{(2\pi)^3}
          \Bigl[ E_{\bm p} \notag\\
         & \hspace{-4mm}+ \frac{T}{N_{\rm c}}
            \ln~ [1 + 3(\Phi+{\bar \Phi} e^{-\beta E_{\bm p}}) 
            e^{-\beta E_{\bm p}}+ e^{-3\beta E_{\bm p}}]
          \notag\\
         & \hspace{-4mm}+ \frac{T}{N_{\rm c}} 
            \ln~ [1 + 3({\bar \Phi}+{\Phi e^{-\beta E_{\bm p}}}) 
               e^{-\beta E_{\bm p}}+ e^{-3\beta E_{\bm p}}]
 	      \Bigl].
 \label{PNJL-Omega} 
 \end{align}
 for $\beta=1/T$, $\Phi={\bar \Phi}$ and $E_{\bm p}=\sqrt{M+{\bm p}^2}$. 

\subsection{$\chi$-PV regularization}
\label{Sec-chi-PV regularization}

In Eq.~\eqref{PNJL-Omega},  the momentum $\bm{p}$ integral has ultraviolet divergence and it must be regularized. 
As an usual regularization scheme, three-dimensional momentum cutoff has been commonly used so far, 
but the regularization explicitly breaks translational invariance 
that is essential for deriving the screening mass~\cite{Ishii:2013kaa}.
Moreover, translational invariance is also necessary to maintain Ward-Takahashi identities for $SU(2)_{\rm V}$ and $SU(2)_{\rm A}$ currents that realize transverse and longitudinal properties of $\rho$ and $A_1$ mesons. 
Hence we take a chiral version~\cite{Ishii:2013kaa}  of Pauli-Villars (PV) regularization~\cite{PV,Florkowski} in the present work. 
We refer to it as $\chi$-PV regularization. 
The original PV regularization cannot maintain chiral symmetry due to heavy masses of auxiliary particles,
but $\chi$-PV regularization of Ref.~\cite{Ishii:2013kaa} manifestly preserves chiral symmetry and correctly reproduces low-energy relations of chiral dynamics, such as the Gell-Mann--Oakes--Renner relation, partial conserved axial-vector current relation and so on. 

In $\chi$-PV regularization, the integral $\Omega_{\rm F}(M)$ is simply regularized as 
\begin{equation}
\Omega_{\rm F}(M)\to\Omega^{\rm reg}_{\rm F}(M)
= \sum_{\alpha=0}^2 C_\alpha \Omega_{\rm F}(M_\alpha) 
\label{PV}
\end{equation}
with $M_0=M$ and the $M_\alpha~(\alpha\ge 1)$ are masses of auxiliary 
particles. The $M_\alpha$ and the $C_\alpha$ 
should satisfy the condition  
$\sum_{\alpha=0}^2C_\alpha=\sum_{\alpha=0}^2 C_\alpha M_\alpha^2=0$ 
to remove the logarithmic, quadratic and quartic divergence. 
We then assume $(C_0,C_1,C_2)=(1,-2,1)$ and $(M_1^2,M_2^2)=(M^2+\Lambda^2,M^2+2\Lambda^2)$. 
The dimensionful parameter $\Lambda$ should be kept to finite 
even after the subtraction \eqref{PV}, 
since the present model is non-renormalizable.

\subsection{Parameter fitting}
\label{Parameter fitting}

In the case of $T=0$~MeV, the present model has four parameters $m_0,G_{\rm S}(0),G_{\rm V}(0)$ 
and the cutoff $\Lambda$. 
We fix the current quark mass $m_0$ to $3.5$ MeV, and determine the $G_{\rm S}(0),G_{\rm V}(0)$, $\Lambda$ 
from three realistic values of $M_\pi=140$ MeV, the pion decay constant $f_\pi= 93.4$ MeV 
and $\rho$ meson mass $M_\rho=770$ MeV; see Table \ref{Model parameters} for the values of 
$m_0, \Lambda, G_{\rm S}(0)\Lambda^2, G_{\rm V}(0)/G_{\rm S}(0)$. 

\begin{table}[h]
\begin{center}
\caption
{Model parameters in the NJL part. }

\begin{tabular}{lcccc}
\hline\hline
$m_0$~[{\rm MeV}]
&$\Lambda$~[{\rm MeV}]
&$G_{\rm S}(0)\Lambda^2$
&$G_{\rm V}(0)/G_{\rm S}(0)$
\\
\hline
3.5
&900
&3.30
&-1.36
\\
\hline
\\
\hline\hline
$T_{\rm S}$~[{\rm MeV}]
&$b_{\rm S}$~[{\rm MeV}]
&$T_0$~[{\rm MeV}]
\\
\hline
135
&115
&215
\\
\hline
\end{tabular}
 \label{Model parameters}
\end{center}
\end{table}

For finite $T$, the present PNJL model with three adjustable parameters, i.e.,  
$(T_{\rm S},b_{\rm S})$ in scalar-type coupling $G_{\rm S}(T)$ and a constant $T_{0}$ 
in  the logarithm-type and the polynomial-type $\mathcal{U}$. 
These parameters are determined so as to reproduce LQCD data 
on $T$ dependence of chiral condensate $\sigma$; 
the chiral- and the deconfinement-transition temperature 
satisfy $T_{\rm c}^{\chi,{\rm LQCD}}=T_{\rm c}^{\rm d,LQCD} = 173$ MeV 
within 10\% errors~\cite{Karsch4}. 
The parameters thus obtained are $T_0=215$~MeV and $(T_{\rm S},b_{\rm S})=(135~{\rm MeV},115~{\rm MeV})$; 
Eventually, all the values shown in Table \ref{Model parameters} are independent of the type of $\mathcal{U}$. 

\begin{figure}[H]
\begin{center}
  \includegraphics[width=0.45\textwidth]{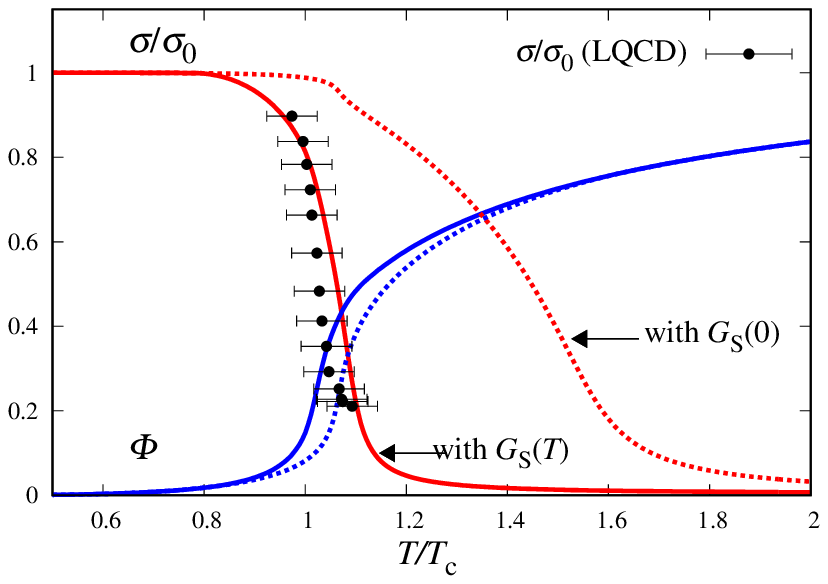}
  \includegraphics[width=0.45\textwidth]{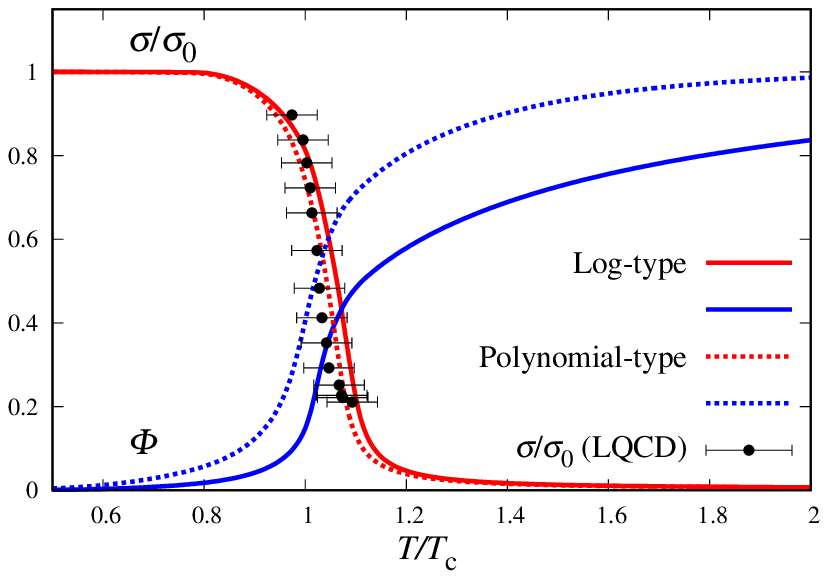}  
\end{center}
\caption{$T$ dependence of the chiral condensate $\sigma$ and 
 the Polyakov loop $\Phi$. The upper panel is results of the logarithm-type  $\mathcal{U}$, 
 while the lower panel is ones of the polynomial-type $\mathcal{U}$.  
The horizontal axis is scaled by $T_{\rm c}^{\chi}$. 
The chiral condensate is normalized by its value $\sigma_0$ at $T=0$. LQCD data are taken from Refs.~\cite{Karsch3,Karsch4}. The 10 \% errors
 come from $T_{\rm c}^{\rm d,LQCD}$ and $T_{\rm c}^{\chi,{\rm LQCD}}$. 
In the upper panel, the solid lines denote the results of the present model with $T$-dependent $G_{\rm S}(T)$, whereas 
the dashed lines correspond to the results of  the present model with a constant $G_{\rm S}(0)$.
} 
\label{fig:condensate}
\end{figure}

Figure~\ref{fig:condensate} shows two cases of the logarithm-type and the polynomial-type $\mathcal{U}$. 
Comparing the solid line with the the corresponding dashed one for $\sigma/\sigma_0$, 
we find that $T$ dependence of $G_{\rm S}(T)$ is essential 
to explain the rapid decrease of $\sigma$ around $T=T_{\rm c}^{\chi,{\rm LQCD}}$~\cite{Karsch3,Karsch4}.

\subsection{Magnetic gluon contribution}
\label{Magnetic gluon contribution}

Now we consider the Polynomial-type $\mathcal{U}$, since  it yields better agreement  with LQCD data on 
${M_{\pi}^{\rm scr}}(T)$ and ${M_{\rho}^{\rm scr}}(T)$ than the logarithm-type one; see Fig.~\ref{pi_rho_Polyakov-0}. 
In the PNJL model, magnetic gluon $\bm{A}^a(x)$ has been usually ignored, 
but the contribution should be important in the deconfinement phase $T>T_{\rm c}^{\rm d}$~\cite{Laine:2003bd}.
If one assumes the meson propagation into $z$-direction for convenience,
the theory tells us that the "electric" gluon field $A_4^{a}$ is totally canceled out by $z$-component of $\bm{A}^a$
and remaining $A_x^{a}$ and $A_y^{a}$ fields induce the positive mass-shift $C_{\rm F}/{8\pi}$ in the quark thermal mass 
 $\mathcal{M}_{\rm thermal}$; namely, 
\beq
\mathcal{M}_{\rm thermal} = \pi T + g_{\rm E}^2
\frac{C_{\rm F}}{8\pi} +\mathcal{O}(g^2_{\rm E}/T)
\eeq
with $C_{F}=(N_{\rm c}^2-1)/2N_{\rm c}$. 
We introduce effects of the shift with the replacement 
\bea
\mathcal{M}_{j,n=0,\alpha}= \sqrt{M_\alpha^2 + (\pi T + T\phi_{j})^2} 
\to \mathcal{M}_{\rm thermal},~~~ 
\\
\mathcal{M}_{j,n=-1,\alpha}= \sqrt{M_\alpha^2 + (-\pi T + T\phi_{j})^2}
\to - \mathcal{M}_{\rm thermal}~~~~~
\label{}
\eea
in $T>T_{\rm c}^{\rm d}$, 
although the other Matsubara frequencies $n$ are untouched.
We have used $T$ dependence of gauge coupling $g_{\rm E}^2$ calculated
with two-flavor up to two-loop order in Ref.~~\cite{Laine:2003bd}, 
where they optimize the scale parameter $\bar{\mu}$ and show 
$g_{\rm E}^2$ is less sensitive to variation around the optimized scale
parameter $\bar{\mu}_{\rm opt}$; 
therefore we choose  $\bar{\mu}=\bar{\mu}_{\rm opt}$ in the present calculation.
The remaining parameter $\Lambda_{\bar{\rm MS}}$ should be related with typical energy scale of QCD. 
The ratio $\Lambda_{\bar{\rm MS}}/T_{\rm c}^{\rm d}$ has been estimated 
from LQCD data on zero temperature string tension or Sommer scale~\cite{Laine:2003bd}, 
and $\Lambda_{\bar{\rm MS}}/T_{\rm c}^{\rm d}$ has been almost equal to 1 within a few 10\% error. 
Hence $\Lambda_{\bar{\rm MS}} \simeq T_{\rm c}^{\rm d}=173$ MeV is simply assumed  here. 

\begin{figure}[htbp]
\begin{center}
  \includegraphics[width=0.45\textwidth]{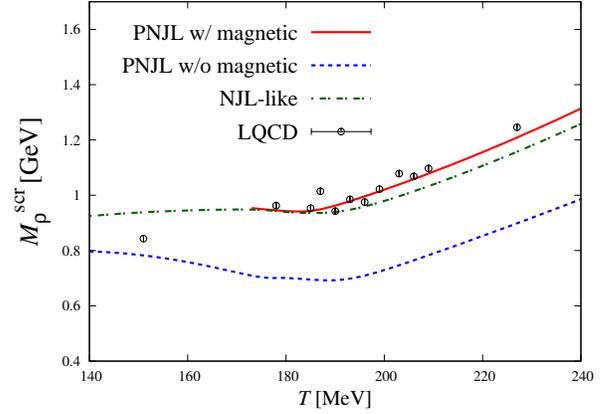}
\end{center}
\caption{$T$ dependence of transverse $\rho$-meson  the Polynomial-type $\mathcal{U}$. 
The solid line stands for the result of MG-PNJL model. 
The dashed line denotes the result of the present PNJL model, while  the dot-dashed line is a result of 
the NJL-like model that is explained in Sec.~\ref{The comparison with LQCD data}.    
} 
\label{rho_NJL_vs_PNJL}
\end{figure}

The PNJL  result  with the replacement is nothing but MG-PNJL model 
The MG-PNJL result (solid line) is shown in  Fig.~\ref{rho_NJL_vs_PNJL} for ${M_{\rho}^{\rm scr}}(T)$. 
The result is valid in $T>T_{\rm c}^{\rm d}$. Effects of the replacement are visible for ${M_{\pi}^{\rm scr}}(T)$.
In a NJL-like model, we use the PNJL model with  $G_{\rm S}(T)$ for $\Omega$ and 
set $\phi_1=\phi_2=\phi_3=0$ in Eqs.~\eqref{SD_rho_A1} and \eqref{SD_pi} for screening masses.

\section{Numerical Results}
\label{Numerical Results}

\subsection{The comparison with LQCD data}
\label{The comparison with LQCD data}

\begin{figure}[H]
\begin{center}
  \includegraphics[width=0.45\textwidth]{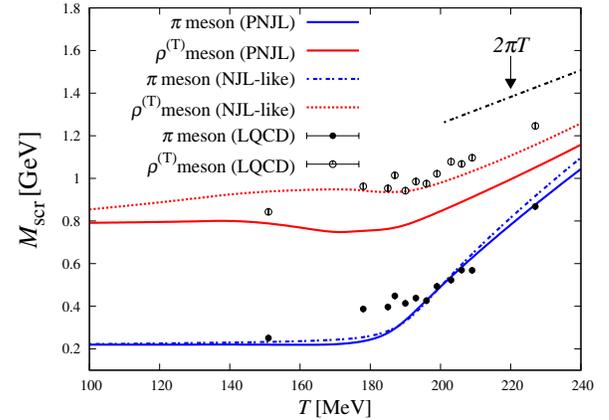}
\end{center}
\caption{$T$ dependence of $\pi$ and transverse $\rho$ mesons. 
with the Polynomial-type $\mathcal{U}$. } 
\label{pi_rho_NJL_vs_PNJL}
\end{figure}

Figure \ref{pi_rho_NJL_vs_PNJL} shows $\pi$ and $\rho$-meson screening masses
calculated with PNJL and NJL-like models. 
For $\pi$ meson, the difference between PNJL and NJL-like results is tiny.
The difference indicates that effects of $\phi_i$ are small for ${M_{\pi}^{\rm scr}}(T)$. 
For $\rho$ meson, the PNJL model agrees with LQCD data in the confinement phase $T < T_{\rm c}^{\rm d}=173$~MeV, 
but not in the deconfinement phases. 
The PNJL model shows  mass reduction around $T\simeq T_{\rm c}^{\chi}$.  
This is attributed to decrease of effective-quark-mass $M$.

For transverse $\rho$ meson, the NJL-like result is above the PNJL one. 
Considering the consistency with LQCD data, 
one find that the PNJL model is more preferable than the NJL-like model 
in the confinement phase $T < T_{\rm c}^{\rm d}=173$~MeV.  
The failure of PNJL model in the deconfiment phase $T > T_{\rm c}^{\rm d}=173$~MeV comes from absence 
of magnetic gluon contribution, as shown in Fig.~\ref{pi_rho_NJL_vs_PNJL}. 

\subsection{$T$ dependent of $G_{\rm V}$}
\label{$T$ dependent of $G_{rm V}$}

\begin{figure}[H]
\begin{center}
  \includegraphics[width=0.45\textwidth]{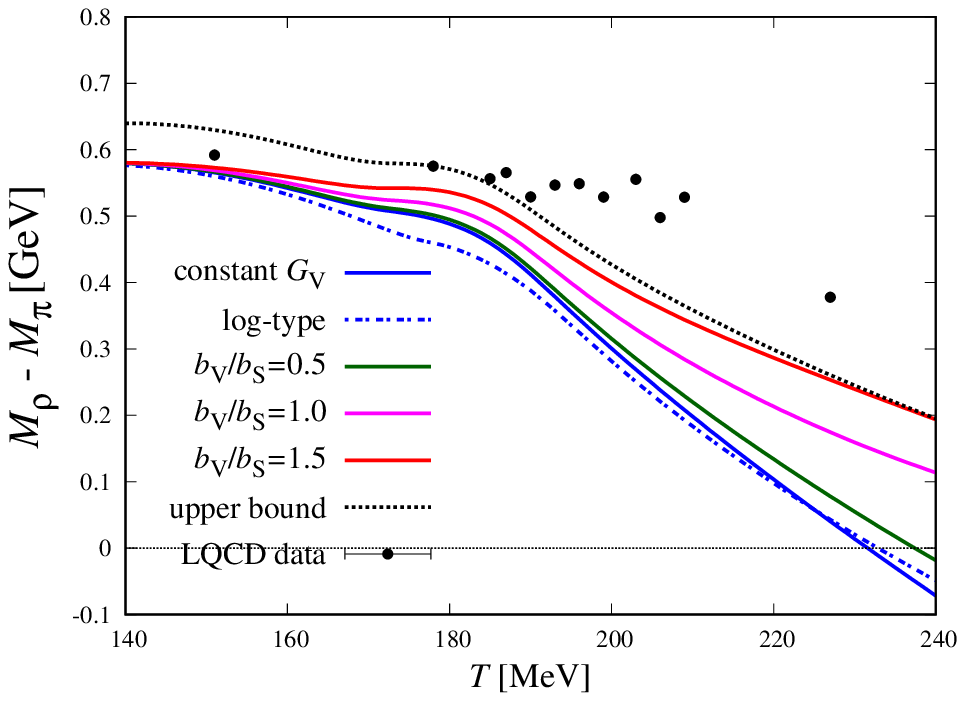}
   \includegraphics[width=0.45\textwidth]{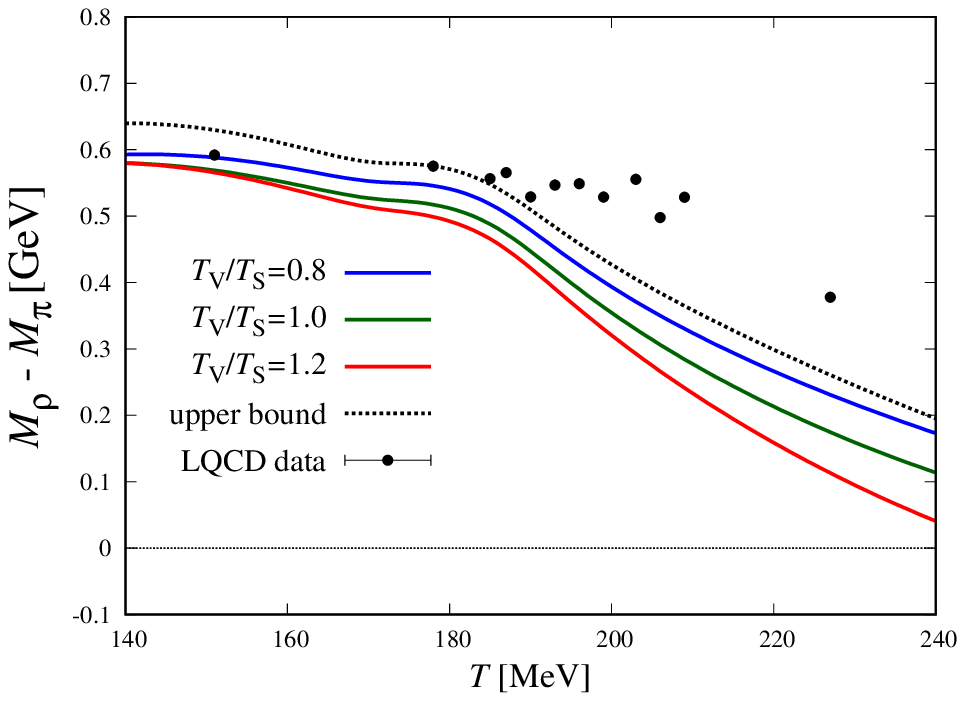}
\end{center}
\caption{Different sets of  $(T_{\rm V},b_{\rm V})$. 
LQCD data are taken from Ref.~\cite{Cheng:2010fe,Maezawa:2016pwo}.
} 
\label{pi_rho_Polyakov_gvT}
\end{figure}

Now we consider  $T$ dependent of $G_{\rm V}(T)$ and assume the same form as $G_{\rm S}(T)$. 
As shown in Fig.~\ref{pi_rho_Polyakov_gvT}, the $(T_{\rm V},b_{\rm V})$ controls $T$ dependence 
of the mass difference between $\pi$ and $\rho$ meson screening masses.

\section{Summary}
\label{Summary} 

We consider the two-flavor thermal system having $\mu=0$, and focus on $\pi$, $\rho$ mesons. 
As a model consistent with LQCD data 
on ${M_{\rho}^{\rm scr}}(T)$ and ${M_{\pi}^{\rm scr}}(T)$, we construct the MG-PNJLmodel. 
Whenever we consider  $\pi$, the mixing between $\pi$ and $A_1$  is taken into account. 

As shown in  the upper panel  of Fig.~\ref{pi_rho_Polyakov-0}, the PNJL model does not reproduce LQCD data~\cite{Cheng:2010fe,Maezawa:2016pwo} on ${M_{\rho}^{\rm scr}}(T)$ 
above $T_{\rm c}^{\chi}\simeq T_{\rm c}^{\rm d} \approx173$ MeV, when we take
the logarithm-type Polyakov-loop potential $\mathcal{U}$ of Ref.~\cite{Rossner}. 
In the lower panel, we take the polynomial-type  $\mathcal{U}$ of Ref.~\cite{Haas:2013qwp}. 
The ${M_{\rho}^{\rm scr}}(T)$ for the polynomial-type  $\mathcal{U}$ is better agreement with the corresponding LQCD data than that for the logarithm-type Polyakov-loop  $\mathcal{U}$. 
We then took the polynomial-type  $\mathcal{U}$. 
The polynomial-type  $\mathcal{U}$ reproduces the LQCD data for ${M_{\pi}^{\rm scr}}(T)$.  
Finally we consider magnetic-gluon contribution on ${M_{\rho}^{\rm scr}}(T)$ and ${M_{\rho}^{\rm scr}}(T)$.  
The results are consistent with the LQCD data for  ${M_{\rho}^{\rm scr}}(T)$ in $T > T_{\rm c}^{\rm d}$ and 
for  ${M_{\pi}^{\rm scr}}(T)$ in both $T > T_{\rm c}^{\rm d}$ and $T < T_{\rm c}^{\rm d}$. 
The present version of PNJL model is referred to as ``magnetic-gluon  (MG) PNJL''  in this paper.

\noindent
\begin{acknowledgments}
The authors thank to Okuto Morikawa for fruitful discussions. 
\end{acknowledgments}

\noindent
\appendix

\section{Meson screening mass}
\label{Meson screening mass}
We derive the equation for $\pi$-, $\rho$- and $A_1$-meson screening masses by using
the method of Refs.~\cite{Ishii:2013kaa}.

The mesonic correlation function corresponding to $\pi,\rho,A_1$ meson is 
\beq
J_{\xi}^{a}(x) = \bar {\psi}(x) \Gamma_\xi \tau^a \psi(x),\nonumber\\
 \eeq
with matrices
\beq
 \pi:~\Gamma_{\rm P} = i\gamma_5,~~
\rho:~\Gamma_{\rm V} = \gamma^\mu,~~
A_1:~\Gamma_{\rm A} = \gamma^{\mu}\gamma_5.
 \label{currents}
\eeq
The mesonic correlation function is defined by 
\begin{align}
\eta_{\xi\xi'}^{ab} (x) \equiv 
\langle 0 | {\rm T} 
\left( J_{\xi}^{a}(x) J_{\xi'}^{b}(0) \right) 
| 0 \rangle 
\end{align}
and their Fourier transformed functions $\chi_{\xi\xi'}^{ab}$ is then obtained by 
\begin{align}
\chi_{\xi\xi'}^{ab} (q^2) 
=  i \int d^4x e^{i q\cdot x}
 \langle 0 | {\rm T} 
\left( J_\xi^a (x) J_{\xi'}^b(0) \right)  
 | 0 \rangle.
\end{align}
where the indices $\xi,\xi'$ means the $\pi,\rho,A_1$ channels as shown in \eqref{currents} and  {\rm T} stands for the time ordered product. 
In the isospin symmetric case , mesonic correlation $\chi_{\xi\xi'}^{ab}$ 
is simplified with $\delta^{ab}\chi_{\xi\xi'}$; therefore, we abbreviate the indices $a,b$ unless otherwise stated.

The ring approximation in $\chi_{\xi\xi'}$ leads to the Schwinger-Dyson equation 
\bea
  \chi_{\xi\xi'}= \Pi_{\xi\xi'}
+ \sum_{\xi''\xi'''}\Pi_{\xi\xi''}2G_{\xi''\xi'''}\chi_{\xi'''\xi'},
  \label{SD-eq}
\eea
where the one-loop polarization function 
$ \Pi_{\xi\xi'}$ is defined as 
\bea
  \Pi_{\xi\xi'} & \equiv & -2i \int \frac{d^4 p}{(2\pi)^4} 
  {\rm tr}_{\rm c,d} \left[\Gamma_{\xi} iS(p'+q) \Gamma_{\xi'}
  iS(p') \right]\nonumber\\
\label{Eq:one-loop polarization function}
\eea
with $p'=(p_{0}+iA_4,{\bm p})$ and trace ${\rm tr}_{\rm c,d}$ of color and Dirac spaces. 
We first consider the zero temperature case for simplicity. 
In this case, the polarization functions are summarized by  
\begin{eqnarray}
\mathit{\Pi}_{\rm PP}
&=&-2i\int\frac{d^{4}p}{(2\pi)^{4}}
{\rm tr}_{\rm c,d}\left[i\gamma_{5}iS(p'+q)i\gamma_{5}iS(p')\right]\\
\mathit{\Pi}_{\rm VV}^{\mu\nu}
&=&-2i\int\frac{d^{4}p}{(2\pi)^{4}}
{\rm tr}_{\rm c,d}\left[\gamma^{\mu}iS(p'+q)\gamma^{\nu}iS(p')\right]\\
\mathit{\Pi}_{\rm PA}^{\mu}
&=&-2i\int\frac{d^{4}p}{(2\pi)^{4}}
{\rm tr}_{\rm c,d}\left[i\gamma_5iS(p'+q)\gamma^{\mu}\gamma_5iS(p')\right]\\
\mathit{\Pi}_{\rm AP}^{\mu}
&=&-2i\int\frac{d^{4}p}{(2\pi)^{4}}
{\rm tr}_{\rm c,d}\left[\gamma^\mu\gamma_5iS(p'+q)i\gamma_5iS(p')\right]\\
\mathit{\Pi}_{\rm AA}^{\mu\nu}
&=&-2i\int\frac{d^{4}p}{(2\pi)^{4}}
{\rm tr}_{\rm c,d}\left[\gamma^{\mu}\gamma_5iS(p'+q)\gamma^{\nu}\gamma_5iS(p')\right].\nonumber\\
\end{eqnarray}   
Taking the trace ${\rm tr}_{\rm d}$ of Dirac index, we obtain the explicit form of polarization functions:  
\begin{eqnarray}
\mathit{\Pi}_{\rm PP}
&=&8I_1 - 4q^2I_2\\
\mathit{\Pi}_{\rm PA}^{\mu}
&=& - \mathit{\Pi}_{\rm AP}^{\mu}
= 8iMq^{\mu}I_2.
\end{eqnarray}   
for psedoscalar and $\pi$-$A_1$ mixing channel, 
where the functions $I_!$ and $I_2$ is represented by 
\bea
I_{1}&=&i{\rm tr_c}\int {d^4p\over{(2\pi )^4}}
{1\over{p'^2-M^2} + i\epsilon},
\label{I1}
\\
I_{2}&=&i{\rm tr_c}\int {d^4p\over{(2\pi )^4}}
{1\over{\{(p'+q)^2-M^2+i\epsilon\}(p'^2-M^2+i\epsilon)}},
\nonumber\\
\label{I2}
\eea
see Appendix for momentum $p$-integrated form of $I_1$ and $I_2$.
In the vector and axial-vector channel, 
$\Pi_{\rm VV}^{\mu\nu}$ and $\Pi_{\rm AA}^{\mu\nu}$ 
are  represented by 
\begin{eqnarray}
\mathit{\Pi}_{\rm VV}^{\mu\nu}
&=&-8i{\rm tr}_{\rm c}
\int\frac{d^{4}p}{(2\pi)^{4}}
\nonumber\\
&&\frac{
2p'^{\mu}
p'^{\nu}
-
q^{\mu}
q^{\nu}/2
-
\left(p'^2 - q^2/4 - M^2\right)g^{\mu\nu}}
{\left\{(p'+q/2)^{2}-M^2 + i\epsilon\right\}
\left\{(p'-q/2)^{2}-M^2 + i\epsilon\right\}},
\nonumber\\
\\
\mathit{\Pi}_{\rm AA}^{\mu\nu}
&=&
\mathit{\Pi}_{\rm VV}^{\mu\nu}
-16M^2I_2 g^{\mu\nu}
\end{eqnarray}

The extension to finite $T$ case can be made by following replacement:
\begin{align}
&p_0 \to i \omega_n = i(2n+1) \pi T, 
\nonumber\\
&\int \frac{d^4p}{(2 \pi)^4} 
\to iT\sum_{n=-\infty}^{\infty} \int \frac{d^3p}{(2 \pi)^3}. 
\label{finte_T_mu}
\end{align}
One can then obtain PV regularized functions $I_1^{\rm reg},I_2^{\rm reg}$ at finite $T$ 
in the static limit as 
\bea
I_{1}^{\rm reg}&=&T\sum_{j,n,\alpha}\int {d^3\bm{p}\over{(2\pi )^3}}
C_\alpha{1\over{\bm{p}^2 + \mathcal{M}_{j,n,\alpha}^2}}
\label{I1_tem}
\\
I_{2}^{\rm reg}&=&-T\sum_{j,n,\alpha}\int {d^3\bm{p}\over{(2\pi )^3}}
C_\alpha{1\over{(\bm{p+q})^2 + \mathcal{M}_{j,n,\alpha}^2}}
{1\over{\bm{p}^2 + \mathcal{M}_{j,n,\alpha}^2}}
\nonumber\\
\label{I2_tem}
\eea
with the summation $\sum_{j,n,\alpha}=\sum_{j=1}^{N_{\rm c}}\sum_{n=-\infty}^{\infty}\sum_{\alpha=0}^{2}$,  
where the summation $\sum_{\alpha=0}^2$ should be taken before the summation $\sum_{n=-\infty}^\infty$ and $\bm{p}$-integral 
for convergence. 
The mass $\mathcal{M}_{j,n,\alpha}$ depends on temperature and phase $\phi_{j}$ as  
\beq
 \mathcal{M}_{j,n,\alpha}
 = \sqrt{M_\alpha^2 + (\omega_n + T\phi_{j})^2}.
\label{Eq:thermal quark mass}
\eeq
We mention $\mathcal{M}_{j,n,\alpha}$ as "thermal quark mass" 
since $\mathcal{M}_{j,n,\alpha}$ acts as (chiral symmetric) quark mass in 3-dimensional momentum $\bm{p}$ space
of Eqs.~\eqref{I1_tem} and \eqref{I2_tem} .

\section{$\rho$ and $A_1$ meson screening masses at finite $T$}

In this section, we only consider the $\xi={\rm V,\rm A}$
channels. At zero temperature, the polarization functions
$\mathit{\Pi}_{\xi\xi}^{\mu\nu}$ and mesonic correlation
$\chi_{\xi\xi}^{\mu\nu}$ are decomposed into 4-dimensionally transverse and longitudinal modes:
\bea
\mathit{\Pi}_{\xi\xi}^{\mu\nu}
&=& \mathit{\Pi}_{\xi\xi}^{({\rm T,4d})} T^{\mu\nu}
+ \mathit{\Pi}_{\xi\xi}^{({\rm L,4d})} L^{\mu\nu}
\nonumber\\
\chi_{\xi\xi}^{\mu\nu}
&=& \chi_{\xi\xi}^{({\rm T,4d})} T^{\mu\nu}
+ \chi_{\xi\xi}^{({\rm L,4d})} L^{\mu\nu}
\eea
The projection tensors $T^{\mu\nu}$ and $L^{\mu\nu}$ are
defined by 
\beq
T^{\mu\nu} = g^{\mu\nu} - \frac{q^\mu q^{\nu}}{q^2},~~
L^{\mu\nu} = \frac{q^\mu q^\nu}{q^2}.
\eeq
It is noted that the longitudinal element $\Pi_{\rm VV}^{({\rm L,4d})}$ 
must vanish due to isospin symmetry 
since the isovector current should be conserved ($\partial_\mu J_{\rm V}^{\mu a}(x)=0$)
and the corresponding Ward-Takahashi identity indicates $q_\mu\Pi_{\rm
VV}^{\mu\nu}=0$ and $q_\mu\chi_{\rm
VV}^{\mu\nu}=0$. The vanishment of $\Pi_{\rm VV}^{({\rm L,4d})}$ and
$\Pi_{\rm VV}^{({\rm L,4d})}$ is realized
even at finite $T$ because the current conservation law holds for any $T$.
If one chooses three-dimensional momentum-cutoff regularization, 
the Ward-Takahashi identity is spoiled and 
the above discussion is no longer valid due to the lack of translational invariance.

At finite $T$, the 4-dimensionally transverse mode is decomposed into
3-dimensionally transverse and longitudinal modes in the polarization
function and mesonic correlation:
\bea
\mathit{\Pi}_{\xi\xi}^{\mu\nu}
&=& \mathit{\Pi}_{\xi\xi}^{({\rm T,3d})} P^{\mu\nu}_{\rm T}
+ \mathit{\Pi}_{\xi\xi}^{({\rm L,3d})} P^{\mu\nu}_{\rm L}
+ \mathit{\Pi}_{\xi\xi}^{({\rm L,4d})} L^{\mu\nu},
\nonumber\\
\chi_{\xi\xi}^{\mu\nu}
&=& \chi_{\xi\xi}^{({\rm T,3d})} P^{\mu\nu}_{\rm T}
+ \chi_{\xi\xi}^{({\rm L,3d})} P^{\mu\nu}_{\rm L}
+ \chi_{\xi\xi}^{({\rm L,4d})} L^{\mu\nu}.
\eea
The 3-dimensional projection tensors are defined by 
\beq
P^{\mu\nu}_{\rm T}
=g^{\mu\nu} -  \frac{q^{\mu}q^{\nu}}{q^2} -  \frac{n^{\mu}n^{\nu}}{n^2}
,~~
P^{\mu\nu}_{\rm L}= \frac{n^{\mu}n^{\nu}}{n^2},
\eeq
where the 4-dimensional vector $n^{\mu}$ is orthogonal component of heat
bath velocity $u^{\mu}$ against the external momentum $q^{\mu}$, i.e.,
$n^{\mu}=T^{\mu\nu}u_{\nu}$. When one takes the rest frame of heat bath
($u^{\mu}=(1,\bm{0})$) and the static limit, the vector $n^{\mu}$
turns out to be equal to $u^{\mu}$. 
The general discussion of projection tensor is summarized in Appendix.

In the SD equation \eqref{SD-eq}, 3-dimensionally transverse and
longitudinal modes has been completely decoupled, and one can
independently solve the equations for each modes. The solution of SD
equation is finally obtained as   
\bea
\chi_{\xi\xi}^{({\rm T,3d})}
=
\frac{\Pi_{\xi\xi}^{({\rm T,3d})}}
{1 - 2G_{\rm V} \Pi_{\xi\xi}^{({\rm T,3d})}}
,~~
\chi_{\xi\xi}^{({\rm L,3d})}
= 
\frac{\Pi_{\xi\xi}^{({\rm L,3d})}}
{1 - 2G_{\rm V} \Pi_{\xi\xi}^{({\rm L,3d})}}
\nonumber\\
\eea
for $\xi={\rm A},{\rm V}$. 
 The polarization functions are simple form 
\bea
\Pi_{\rm VV}^{({\rm T,3d})} 
&=& -16\bm{q}^2 I_3^{\rm reg},~~
\Pi_{\rm AA}^{({\rm T,3d})}
=-16\bm{q}^2 I_3^{\rm reg} -16M^2 I_2^{\rm reg},
\nonumber \\
\Pi_{\rm VV}^{({\rm L,3d})}
&=& -16I_4^{\rm reg},~~
\Pi_{\rm AA}^{({\rm L,3d})}
=-16I_4^{\rm reg} -16M^2 I_2^{\rm reg},
\eea
where  the functions $I_3^{\rm reg}$ and $I_4^{\rm reg}$ are
\bea
I_{3}^{\rm reg}&=&-T\sum_{j,n,\alpha}
\int {d^3\bm{p}\over{(2\pi)^3}}
\int_{0}^{1} dx~C_\alpha
{x-x^2\over{\bm{p}^2 + \mathcal{M}^2_{j,n,\alpha}(\bm{q}^2)}}
\nonumber\\
\label{I3_tem}
\\
I_{4}^{\rm reg}
&=&\bm{q}^2I_{3}^{\rm reg} 
-T\sum_{j,n,\alpha}
\int {d^3\bm{p}\over{(2\pi)^3}}
\int_{0}^{1} dx~C_\alpha
{\omega_n^{'2} - \bm{p}^2/3
\over{\bm{p}^2 + \mathcal{M}^2_{j,n,\alpha}(\bm{q}^2)}}
\nonumber\\
\label{I4_tem}
\eea
with $\omega_n' = \omega_n + \phi_{j}$, Feynman parameter $x$ and $\mathcal{M}^2_{j,n,\alpha}(\bm{q}^2)
=\mathcal{M}^2_{j,n,\alpha} + (x-x^2)\bm{q}^2$. 
It is worth noting that the Feynman parameter must
be carefully taken in thermal field theory, 
since the naive treatment of Feynman parameter leads to wrong results 
due to ambiguity of the analytic continuation from the Euclid space to the Minkowski space. 
In the present case, the all calculation has been done in the Euclid space 
and such a difficulty does not arise.

The $\rho$- and $A_1$-meson screening masses 
are obtained by searching the pole position of the denominator: 
\bea
\left.
\left[{1 - 2G_{\rm V} \Pi_{\xi\xi}^{({\rm T,3d})}}
\right]\right|_{\tilde{q}=iM_{\rho}^{\rm scr}~{\rm or}~iM_{A_1}^{\rm scr}}
&=&0
\\
\left.
\left[{1 - 2G_{\rm V} \Pi_{\xi\xi}^{({\rm L,3d})}}
\right]\right|_{\tilde{q}=iM_{\rho}^{\rm scr}~{\rm or}~iM_{A_1}^{\rm scr}}
&=&0
\label{SD_rho_A1}
\eea
with $\tilde{q}=|\bm{q}|$. 

\subsection{$\pi$ meson screening mass with $\pi-A_1$ mixing}

The $\pi$ meson is coupled with 4-dimensionally longitudinal mode of $A_1$ meson; 
therefore, the SD equation becomes coupled channel equation: 
\begin{eqnarray}
\chi_{\rm PP}
&=& 
\mathit{\Pi}_{\rm PP}
+ \mathit{\Pi}_{\rm PP} 2G_{\rm S} \chi_{\rm PP}
+ \bar{\mathit{\Pi}}_{\rm PA}2G_{\rm V} \bar{\chi}_{\rm AP}\\
\bar{\chi}_{\rm AP}
&=& 
\bar{\mathit{\Pi}}_{\rm AP}
+ \bar{\mathit{\Pi}}_{\rm AP} 2G_{\rm S} \chi_{\rm PP}
+ \mathit{\Pi}_{\rm AA}^{({\rm L,4d})}2G_{\rm V} 
\bar{\chi}_{\rm AP}\\
\bar{\chi}_{\rm PA}
&=& 
\bar{\mathit{\Pi}}_{\rm PA}
+ \bar{\mathit{\Pi}}_{\rm PA}2G_{\rm V} \chi_{\rm AA}^{({\rm L,4d})}
+ \mathit{\Pi}_{\rm PP}2G_{\rm S} \bar{\chi}_{\rm PA}
\\
\chi_{\rm AA}^{({\rm L,4d})}
&=& 
\mathit{\Pi}_{\rm AA}^{({\rm L,4d})}
+ 
\mathit{\Pi}_{\rm AA}^{({\rm L,4d})}
2G_{\rm V} \chi_{\rm
AA}^{({\rm L,4d})}
+ \bar{\mathit{\Pi}}_{\rm AP}2G_{\rm S} 
\bar{\chi}_{\rm PA}
\nonumber\\
\end{eqnarray}
For convenience, the pseudoscalar-axialvector mixing channels
$\chi_{\rm AP}^{\mu},\chi_{\rm PA}^{\mu},\Pi_{\rm AP}^{\mu},\Pi_{\rm PA}^{\mu}$ 
are rewritten by $\bar{\chi}_{\rm AP(PA)} = \hat{q}_\mu\chi_{\rm
AP(PA)}^{\mu},\bar{\Pi}_{\rm AP(PA)} = \hat{q}_\mu\Pi_{(\rm AP)PA}^{\mu}$ with unit vector $\hat{q}^{\mu}=q^{\mu}/\sqrt{q^2}$. 
When one introduces the following matrices,
\begin{eqnarray}
\boldsymbol{\chi} 
&=& \left(
\begin{array}{cc}
\chi_{\rm PP} & 
\bar{\chi}_{\rm PA} 
\\
\bar{\chi}_{\rm AP} & 
\chi_{\rm AA}^{({\rm L,4d})} 
\end{array}
\right),~~
\bm{G} = 
\left(
\begin{array}{cc}
G_{\rm S}& 
0
\\
0& 
G_{\rm V}
\end{array}
\right)
\nonumber\\
\boldsymbol{\mathit{\Pi}} 
&=&\left( 
\begin{array}{cc}
\mathit{\Pi}_{\rm PP} & 
\bar{\mathit{\Pi}}_{\rm PA} 
\\
\bar{\mathit{\Pi}}_{\rm AP} & 
\mathit{\Pi}_{\rm AA}^{({\rm L,4d})} 
\end{array}
\right),
\end{eqnarray}
the solution of SD equation is easily found as
\begin{eqnarray}
\boldsymbol{\chi} 
= 
\frac
{\tilde{\boldsymbol{\mathit{\Pi}}}\boldsymbol{\mathit{\Pi}}}
{\det{\left[
\bm{I} - 2\boldsymbol{\mathit{\Pi}}\bm{G}
\right]}}.
\end{eqnarray}
The matrix $\tilde{\boldsymbol{\mathit{\Pi}}}$ is 
\begin{equation}
 \tilde{\boldsymbol{\mathit{\Pi}}}
= 
\left(
\begin{array}{cc}
1-2G_{\rm V}\mathit{\Pi}_{\rm AA}^{({\rm L,4d})} &2G_{\rm V}\bar{\mathit{\Pi}}_{\rm PA}\\
2G_{\rm S}\bar{\mathit{\Pi}}_{\rm AP}&1-2G_{\rm S}\mathit{\Pi}_{\rm PP}
\end{array}
\right)
\end{equation}
and the determinant is
\begin{eqnarray}
 \det{
\left[
\bm{I} - 2\boldsymbol{\mathit{\Pi}}\bm{G}
\right]}
&=& 
\left(1 - 2G_{\rm S}\mathit{\Pi}_{\rm PP}\right)
\left(1 - 2G_{\rm V}\mathit{\Pi}_{\rm AA}^{({\rm L,4d})}\right)
\nonumber\\
&-& 4G_{\rm S}G_{\rm V}\bar{\mathit{\Pi}}_{\rm AP}
\bar{\mathit{\Pi}}_{\rm PA}.
\end{eqnarray}
The $\pi$ meson screening mass is then obtained by
\begin{eqnarray}
 \left.
 \det{
\left[
\bm{I} - 2\boldsymbol{\mathit{\Pi}}\bm{G}
\right]}
\right|_{\tilde{q}=iM_{\xi}^{\rm scr}}
=0
\label{SD_pi}
\end{eqnarray}
for $\xi=\pi$ meson and $A_1$ meson in 4-dimensionally longitudinal
state. We numerically found that any pole since . 

\section{Threshold mass}
\label{sec:Threshold mass}

In PNJL model, mesons can decay into quark-pair. 
Such a effect is emerged in the functions $I_2,I_3,I_4$:
\bea
I_{2}^{\rm reg}
&=&
\frac{iT}{8\pi\tilde{q}}{\rm Log}
{\left(\frac{2 \mathcal{M}_{j,n,\alpha}+ i\tilde{q}}
{2 \mathcal{M}_{j,n,\alpha}-i\tilde{q}}\right)}
\nonumber\\
I_{3}^{\rm reg}
&=&
\frac{iT}{64\pi\tilde{q}}
\sum_{j,n,\alpha}
\left[
\left(1- \frac{4\mathcal{M}_{j,n,\alpha}^2}{\tilde{q}^2}\right)
{\rm Log}{\left(\frac{2 \mathcal{M}_{j,n,\alpha}+ i\tilde{q}}
{2 \mathcal{M}_{j,n,\alpha}-i\tilde{q}}\right)}
\right.\nonumber\\
&+& \left.\frac{4i\mathcal{M}_{j,n,\alpha}}{\tilde{q}}\right]
\nonumber\\
\\
I_{4}^{\rm reg}
&=&
\frac{i\tilde{q}T}{32\pi}
\sum_{j,n,\alpha}
\left[
\left(1 + \frac{4\omega_n^2}{\tilde{q}^2}\right)
{\rm Log}{\left(\frac{2 \mathcal{M}_{j,n,\alpha}+ i\tilde{q}}
{2 \mathcal{M}_{j,n,\alpha}-i\tilde{q}}\right)}
\right.\nonumber\\
&+& \left.\frac{4i\mathcal{M}_{j,n,\alpha}}{\tilde{q}}\right],
\eea
as logarithmic cuts along the imaginary axis in complex $\tilde{q}$ plane, 
where logarithmic cuts are starting at $\tilde{q}=2i\mathcal{M}_{j,n,\alpha}$ 
and lowest branch point is given by $\tilde{q}=i\mathcal{M}_{\rm lowest}$ 
with 
\beq
\mathcal{M}_{\rm lowest} = 2\mathcal{M}_{j=1,n=0,\alpha=0}
= 2\mathcal{M}_{j=2,n=-1,\alpha=0}.
\label{threshold}
\eeq
When $\tilde{q} \ge i\mathcal{M}_{\rm lowest}$,  meson decays into quark-pair.
The $\mathcal{M}_{\rm lowest}$ is thus "threshold mass".  

PNJL model describes statistical confinement by means of Polyakov loop, 
but has no information about the confinement force between quarks 
because the gauge fields are treated as background field 
and their nonlocal correlations are ignored. 
Accordingly, PNJL model may be less predictive to meson screening mass above the threshold mass.
Hence we assume that our model results are reliable only when the following relation is satisfied:
\beq
M_{\xi}^{\rm scr}<\mathcal{M}_{\rm lowest}
\label{Eq:threshold condition}
\eeq
for $\xi=\pi,\rho,A_1$ mesons.


\end{document}